\documentclass[doublecol]{epl2}
\usepackage{graphicx}

\def\lsim{\mathrel{\rlap{\lower4pt\hbox{\hskip1pt$\sim$}}
    \raise1pt\hbox{$<$}}}                % less than or approx. symbol
\def\gsim{\mathrel{\rlap{\lower4pt\hbox{\hskip1pt$\sim$}}
    \raise1pt\hbox{$>$}}}                % greater than or approx. symbol

\title{Generation of Kerr non-Gaussian motional states of trapped ions}

\author{
  M. Stobi\'nska\inst{1,2}
  \and A. S. Villar\inst{2,3}\thanks{E-mail: \email{alessandro.villar@mpl.mpg.de}}
  \and G. Leuchs\inst{2,3}}
\shortauthor{M. Stobi\'nska \etal}

\institute{
  \inst{1} Institute for Theoretical Physics II,
  University of Erlangen-Nuremberg, 
  91058 Erlangen, Germany\\
  \inst{2} Max Planck Institute for the Science of Light,
  91058 Erlangen, Germany\\
  \inst{3} Institute of Optics, Information and Photonics,
  University of Erlangen-Nuremberg, 
  91058 Erlangen, Germany
}


\abstract{Non-Gaussian states represent a powerful resource for quantum
information protocols in the continuous variables regime.  
Cat states, in particular, have been produced in the motional 
degree of freedom of trapped ions by controlled displacements
dependent on the ionic internal state. An alternative method 
harnesses the Kerr nonlinearity naturally present in this 
kind of system. We perform detailed calculations confirming 
its feasibility for typical experimental conditions. 
Additionally, this method permits generation of all other complex 
non-Gaussian states with negative Wigner functions resulting from 
Kerr nonlinear interaction. Especially, superpositions of 
several coherent states are achieved at 
a fraction of the time necessary to produce the cat state.}

\begin{document}

\maketitle

\section{Introduction}

Quantum information processing relying on physical observables with continuous
spectra (continuous variables, or CV) offer new possibilities of
implementation and employment when compared to their discrete alternatives.
In particular, states with non-Gaussian Wigner function are
essential for the speedup and universality of CV quantum 
computation~\cite{Lloyd1999,Bartlett2002,grangierkitten}. According to the  
Gottesman-Knill theorem, quantum computing based only on components described 
by quadratic Hamiltonians, Gaussian inputs, and measurements on canonical variables 
can be simulated by a classical computation~\cite{gottesmanknill}. Thus, non-Gaussian states allow to 
exploit the advantage of the quantum algorithms over the classical ones~\cite{Bartlett2002}.
Construction of a CV universal quantum computer for 
transformations that are polynomial in those variables requires at least a cubic (Kerr)
nonlinear operations~\cite{Lloyd1999}. Non-Gaussian states also find important 
applications in high precision metrology~\cite{caves}, novel tests of Bell-like 
inequalities~\cite{carmichaelbell,grangierbell}, and fundamental 
investigations~\cite{Brune1996, Myatt2000,subplanck}. Therefore, finding the effective 
and deterministic way of generation and manipulation of non-Gaussian states and interactions 
is crucial for further development of quantum based applications as well as for
better understanding of physics.

Position and momentum of trapped ions represent a good CV quantum
system candidate, especially taking into account the recent achievements regarding low heating rates~\cite{Chuang}. 
So far, this system is much less exploited than the ionic discrete 
variables~\cite{winelandmotion,winelandcat}. In fact, motional degrees of freedom have been 
mostly used as qubits, fulfilling the role of a quantum information `bus' among 
the ionic internal states~\cite{blattcnot}. However, recently proof-of-principle quantum
walks using the quadratures of the harmonic oscillator state were performed~\cite{Schmitz2009,Zahringer2010} 
and CV quantum simulations were demonstrated~\cite{Gerritsma2010}.%~\cite{Leibfried2002,Gerritsma2010}. 

The current ion trap technology allows an arbitrary quantum manipulation 
of the motional states~\cite{Kneer1998,Gardiner1997,Serra2001}. 
In practice, however, the fidelity of operations changing
the phonon number stepwise (e.g. blue sideband pulses) are comparatively 
low, limiting the maximum achievable phonon number to just a few
quanta in a coherent superposition~\cite{blattmulti}. 
Alternatively, `single-shot' methods relying on a single laser pulse to
directly produce the desired final state seem to
have an advantage from the experimental point of view.
For instance, bichromatic quantum gates allow for high
fidelities in the generation of two-qubit entangled states based on
precise manipulation of the vibrational degree of
freedom~\cite{msgate1,msgate2,rooshighfid}, as well as coherent superpositions 
involving tens of phonons~\cite{Schmitz2009,Zahringer2010}. 

In this paper we investigate a deterministic method to directly harness the Kerr nonlinearity 
present in the motion of trapped ions and thus to produce non-Gaussian vibrational 
states~\cite{davidovichkerr}. The method by employing a single laser pulse allows for creation 
of several types of `quasi macroscopic' coherent state superpositions, such as the cat states, 
and all other non-Gaussian states emerging from Kerr nonlinear interaction. While operations 
capable of producing such states exist in principle, we believe the simplicity of our scheme 
will further push the border of what can be experimentally realized with high fidelities in CV systems.
%Moreover, usually the systems involving the higher-order nonlinearities involve also the lowest 
%one as the strongest component.

This paper is organized as follows. First we introduce the Kerr
non-Gaussian state. In the next section we present the method of
direct generation of this state on a vibrational mode of a trapped
ion.  Then the ideal Kerr state is compared to the result of the
calculation using the complete Hamiltonian of the trapped ions.  We
finish the paper with our conclusions.

\section{The Kerr state}
\label{seckerr}

The lowest order Hamiltonian in powers of the
harmonic oscillator annihilation $a$ and creation $a^\dag$ operators
capable of producing non-Gaussian states describes a Kerr
nonlinear medium.  The evolution of an initial coherent state
$|\alpha\rangle$ subject to this Hamiltonian gives rise to the
so-called `Kerr states',
\begin{equation}
|\Psi(\alpha, \tau)\rangle = e^{-\frac{|\alpha|^2}{2}}
\sum_{n=0}^{\infty}\, \frac{\alpha^n}{\sqrt{n!}}\,
e^{i\frac{\tau}{2}n(n-1)} |n\rangle.
\label{eq:kerr}
\end{equation}
They are parametrized in terms of only one quantity, the effective
time $\tau$~\cite{Tanas}. 
In general, the Kerr states are highly nonclassical and for certain
values of $\tau$ their Wigner function assumes negative
values~\cite{PRA}.  Their most prominent examples are the cat states,
formed by superpositions of two coherent states shifted in phase by
$\pi$.  For instance, the cat state $e^{-i\pi/4}|i\alpha\rangle + e^{i\pi/4}
|-i\alpha\rangle$ results from Eq.~(\ref{eq:kerr}) taken for
$\tau=\pi$.  After this time the evolution effectively reverses to
reach the original coherent state at $\tau=2\pi$. Cat states were
used in studies of decoherence mechanisms and the quantum-classical
boundary~\cite{Brune1996,Myatt2000}.  The `large cats' for which
the two components $|i \alpha\rangle$ and $|-i \alpha\rangle$ are
nearly orthogonal ($\alpha \!>\!1.5$) also find applications in quantum
information processing~\cite{Jeong,Ralph2003}.  Efforts were
made to produce such states using the Kerr nonlinearity in optical
fibers~\cite{leuchsfiber}.  However, optical nonlinearities 
are far too small to enter the highly nonlinear regime and thus to 
obtain nonclassical Wigner functions before dissipation effects 
destroy the coherences.
%~\cite{GJM-CAH}. 
Superpositions of multiple coherent states are obtained for 
$\tau=\pi/2n$, where $n$ is the number of coherent states in the superposition~\cite{Miran1,Miran2}. 
For instance, at $\tau_{\mathrm{eff}} =\pi/2$ one achieves the 
superposition of four of them (the compass
state) and for $\tau_{\mathrm{eff}} = \pi/3$ six coherent states are
superposed~\cite{OSID}.  The maximum number of participating coherent
states relies solely on the physical possibility of distinguishing
them, which depends on the amplitude of the initial coherent
state. Larger values of $\alpha$ will allow more complex
superpositions to be constructed, but will have support on a larger
number of Fock states as well. Such examples of Kerr states
were not experimentally investigated so far.

\begin{figure}
\centering
\includegraphics[scale=0.45]{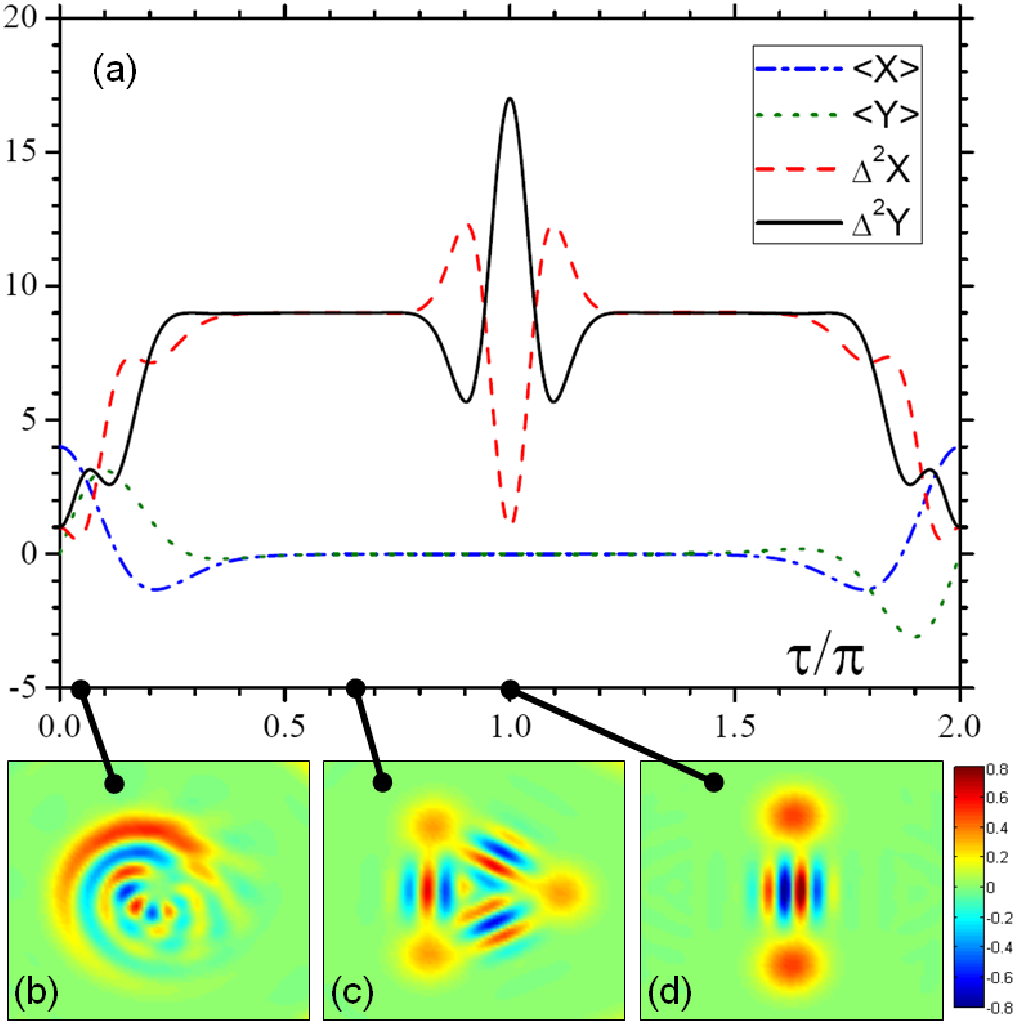}
\caption{Evolution of the ideal Kerr state. (a) Averages and variances
  of the quantum oscillator position $X$ and momentum $Y$ operators as
  a function of the effective time $\tau$~\cite{Tanas}. The insets below show a few
  Wigner functions for specific times~\cite{PRA}: (b) Initial spreading in phase
  space, $\tau=\pi/5$; (c) Superposition of three coherent states,
  $\tau=2\pi/3$; (d) Cat state, $\tau=\pi$.}
\label{kerrideal}
\end{figure}

Time evolution of mean values and variances of the position $X=a+a^\dag$ 
and momentum $Y=-i(a-a^\dag)$ operators in Kerr medium is depicted in 
fig.~\ref{kerrideal}a~\cite{Miran3,Tanas}. The evolution spreads the initial coherent 
state ($\Delta^2 X =\Delta^2 Y =1$)
around the origin of phase space, soon producing negative values in the 
Wigner function. Negative values with amplitude around 10\% of the maximum 
positive value are already observed for $\tau\approx\pi/20$. 
In fig.~\ref{kerrideal}b they reach nearly the same magnitude 
in sub-Planck areas of phase space for $\tau\approx\pi/5$. Since 
the Kerr interaction does not couple different Fock states, 
but instead only dephases them independently, the Wigner function 
cannot occupy higher phonon numbers than originally present. 
For intermediate times, fig.~\ref{kerrideal}c shows one possible 
superposition of multiple coherent states, forming in this case 
a triangular pattern. The cat state is obtained at the turning point 
of the evolution (fig.~\ref{kerrideal}d), after which the states 
produced appear once more in  reversed order.

\section{Generation of motional non-Gaussian Kerr states} 
\label{sechamiltonian}

Let us consider a single ion trapped in the effective harmonic
potential of a Paul trap interacting with a resonant laser field. By
proper choice of geometry, only one of the three existing vibrational
modes can be made relevant to the dynamics. In this case the
interaction Hamiltonian reads~\cite{Leibfried2003}
\begin{equation}
\label{hamiltonian}
H = \hbar \frac{\Omega}{2} \left\{\sigma^+ % Omega/2 -> Omega
%H = \hbar \frac{\Omega}{2} \left\{\sigma^+ % original
e^{i\eta(ae^{-i\omega t}+a^{\dagger }e^{i\omega t})} + \mathrm{h.c.}\right\},
\end{equation}
where $\Omega$ is the Rabi frequency, $\sigma^+$ is the electronic
state rising operator, 
%$a$ is the vibrational state annihilation operator, 
$\omega$ is the quantum oscillator harmonic frequency, 
and $\eta$ is the Lamb-Dicke parameter. We will further
assume the Lamb-Dicke regime and thus $\langle a^\dag
a\rangle\eta\lsim 1$.
% and $\eta\ll1$. 
%It corresponds to the Lamb-Dicke
%regime where the atomic wavefunction does not spread much compared to
%the wavelength of the driving field. 
It is justified in this case to
expand the exponentials of eq.~(\ref{hamiltonian}) in $\eta$ and
disregard the higher order terms. The Kerr effect is described by the 
term $(a^\dag a)^2$. It dephases each Fock state proportionally 
to its eigenvalue (self-phase modulation)~\cite{davidovichkerr}. 
Keeping only the terms resonant to the atomic transition (carrier) 
up to the forth order in $\eta$, we obtain in the interaction picture
\begin{equation}
H^{\mathrm{res}}_{\mathrm{int}} = \hbar \frac{\Omega}{2} \left\{ 
1\!-\!\frac{\eta^2}{2} \!+\! \frac{\eta^4}{8} \!+\! \left(
\!-\eta^2 \!+\!  \frac{\eta^4}{2}\right)a^{\dagger}a
\!+\!\frac{\eta^4}{4}{a^{\dagger}}^2a^2 \right\}\sigma_x.
\label{heff}
\end{equation}
The first three terms describe the corrected internal transition Rabi
frequency, and do not influence the motional state.  The terms
proportional to $a^\dag a$ represent a rotation in phase space and
can be omitted by considering another rotating frame.  Therefore,
the lowest order term to nontrivially influence the motional
dynamics is indeed the Kerr self-phase modulation. Higher order terms,
although contributing as well, are much smaller by the factor of 
order of $\eta^2$. Similarly, off-resonant terms will also hinder the
ideal dynamics. All these effects are disregarded in the present qualitative discussion, 
but are fully taken into account in the numerical
calculation presented in the next section.

The inclusion of additional vibrational modes brings other 
nonlinear effects such as cross-phase modulation~\cite{roospra}. 
Here we will disregard those interesting but more complicated effects 
by supposing that the Lamb-Dicke parameters of
such modes are much smaller than for the main mode considered.
If more ions participate in the dynamics, all the vibrational 
modes in a given direction must be included for consistency, 
in which case cross-phase modulation is unavoidable. 

The Hamiltonian of eq.~(\ref{heff}) couples the internal and the
external degrees of freedom and results in an internal state-dependent
evolution of the motion. To remove this undesired effect, the atomic
internal state must be initially prepared in an eigenstate of
$\sigma_x$.  Usual $\pi/2$ Rabi pulse with $\pi/2$ phase with 
respect to the main Kerr pulse accomplishes the atomic state
superposition on the time scale of $\Omega^{-1}$, i.e. much faster than
the Kerr evolution time scale $\tau_{\mathrm{eff}}$
of eq.~(\ref{taueff}). Thus, it has a
negligible effect on the ion motional state. With this procedure,
the internal atomic state is separable from the motional state.
The evolution operator acting on the vibrational mode then equals
\begin{equation}
%U(t) = e^{i \frac{1}{4}\Omega\eta^4 {a^{\dagger}}^2a^2 t}. % Omega/2 -> Omega
U(t) = \exp\left\{i \frac{1}{8}\Omega\eta^4 {a^{\dagger}}^2a^2 t\right\}. % original
%U(t) = e^{i \frac{\Omega \eta^4 t}{8} {a^{\dagger}}^2a^2 \. (\sigma^+ 
%  + \sigma^-)}.
  \label{unitarynonlinear}
\end{equation}
From this expression we identify the effective value of the Kerr evolution parameter
\begin{equation}
\label{taueff}
%\tau_{\mathrm{eff}} = \frac{1}{2}\eta^4 \Omega t. % Omega/2 -> Omega
\tau_{\mathrm{eff}} = \frac{1}{4}\eta^4 \Omega t. % original
\end{equation}
The highly non-Gaussian Kerr states require a coherent state as the 
starting point of the nonlinear evolution. It can be prepared
in a number of ways~\cite{winelandmotion,winelandcat}. Once
prepared, any state can be reconstructed by a standard technique using 
displacements and parity measurements~\cite{davidovichwigner}.

A special feature of the method under discussion is the independence
between the cat state `size' (with respect to $\langle a^\dag a 
\rangle=|\alpha|^2$) and the time necessary to produce it. Since
the Kerr evolution recurrence time does not depend on the initial
state, a `small cat' and a `large cat' are generated by the 
same light pulse. This is in sharp contrast to approximative methods 
where the desired final superposition is composed of several laser 
pulses each rising the phonon number by one unity~\cite{Kneer1998}. 
In the scheme we investigate here, the maximum cat size is only 
limited by the trap anharmonicities and decoherence (as  all other methods 
ultimately are). Furthermore, no 
additional truncation at a maximum phonon occupation number is necessary.

A simple estimate using eq.~(\ref{taueff}) shows the feasibility of 
the scheme within the typical experimental conditions. 
We take $\omega=2\pi\times3$~MHz as the quantum oscillator frequency. 
%a value which could be larger in a trap optimized to manipulate vibrational states. 
The Rabi frequency $\Omega=2\pi \times 200$~kHz would result
in off-resonant excitation on the order of 0.1\%. For these
parameters, $\eta=0.1$ would allow achieving the cat state in
$t_\mathrm{cat}\approx 100$~ms. This value lies on the boundary of the
expected motional coherence time, typically less than $1$~ms for small 
traps and above $100$~ms for larger traps. However, because 
of the strong dependence of $t_\mathrm{cat}$ on
the Lamb-Dicke parameter, $\eta=0.2$ would already result in
a pulse duration of only $t_\mathrm{cat}\approx 6$~ms.  In any case, other 
complex Kerr states would be accessible for even shorter pulses, as 
previously discussed, lying well within the expected motional 
coherence time. The Lamb-Dicke parameter can be manipulated to a 
certain extent by changing the oscillator frequency as well as by small 
adjustments in the angle between the oscillator axis and the laser 
propagation vector (or two lasers, in the case of Raman excitation)~\cite{wineland}.

\section{Fidelity of the Kerr state produced by the complete Hamiltonian}
\label{secnumerics}

The complete internal and external dynamics of the trapped ion using the full 
Hamiltonian of eq.~(\ref{hamiltonian}) was computed. The effects of the higher-order 
and off-resonant terms were included in the calculation. As the initial condition for the 
ion motion a coherent state of amplitude $\alpha=2$ was taken, corresponding to 
mean phonon number $\langle a^\dag a\rangle=4$. The oscillator frequency 
$\omega=2\pi\times 3$~MHz was assumed. 

For the numerical calculation the Fock basis was truncated at 
$|n=21\rangle$ in order to obtain the probability smaller than $10^{-19}$ for 
the population of the higher Fock states. The initial electronic state 
$(|g\rangle+|e\rangle)/\sqrt{2}$ was chosen as the eigenstate of $\sigma_x$ 
(where $|g\rangle$ and $|e\rangle$ stand for the ground and the excited
internal state, respectively). The evolution of the vibrational state 
was calculated by applying the evolution operator based on 
the Hamiltonian of eq.~(\ref{hamiltonian}) in steps of $dt=10^{-5}$~ms
or smaller. The states obtained in this manner were compared to 
the ideal Kerr state of eq.~(\ref{eq:kerr}) using the fidelity 
defined as the square of their scalar product.

\begin{figure}
\begin{center}
  \raisebox{3.5cm}{\small (a)}\kern.5cm\includegraphics[width=7cm]{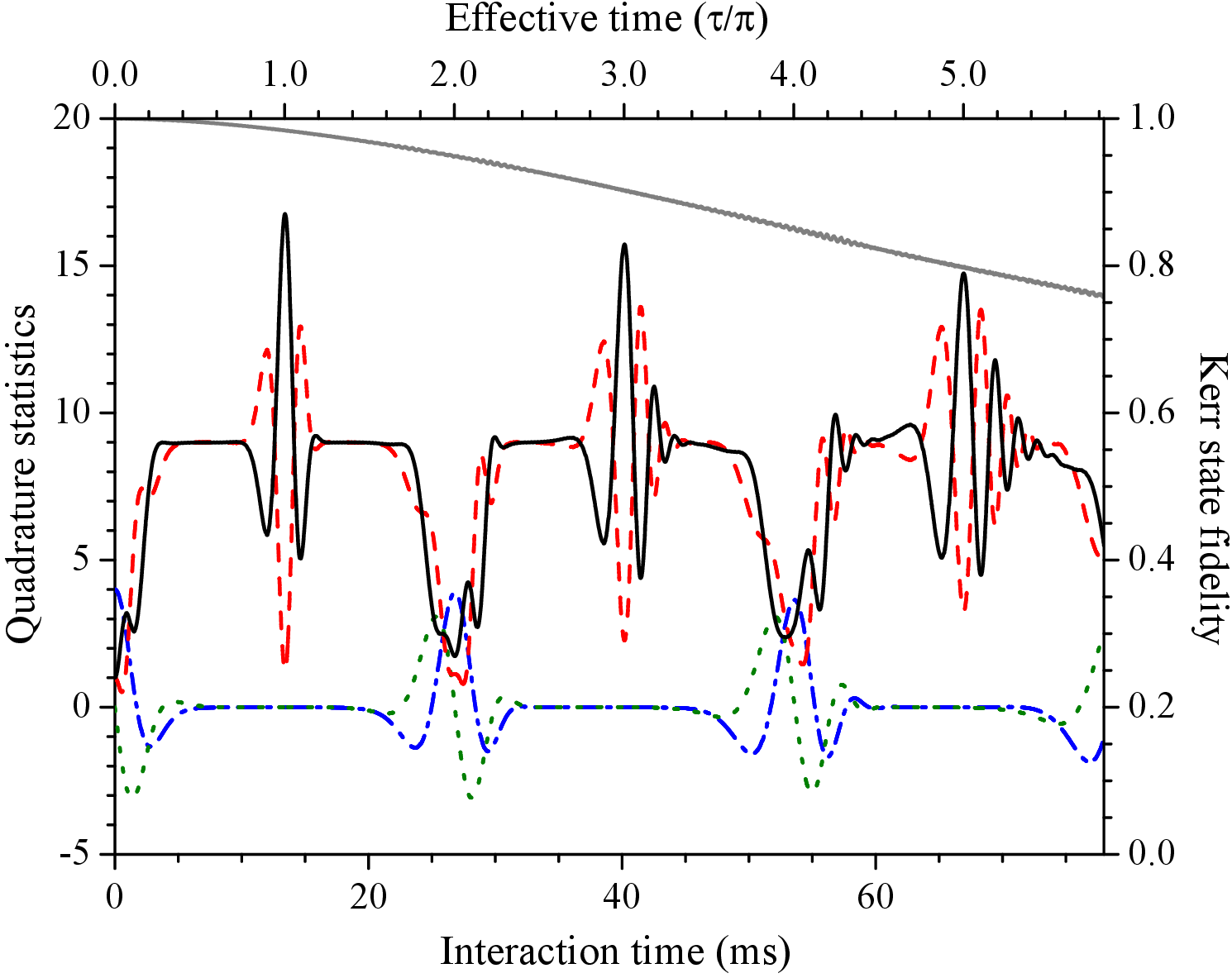}\\[1ex]
  \raisebox{3.5cm}{\small (b)}\includegraphics[width=7cm]{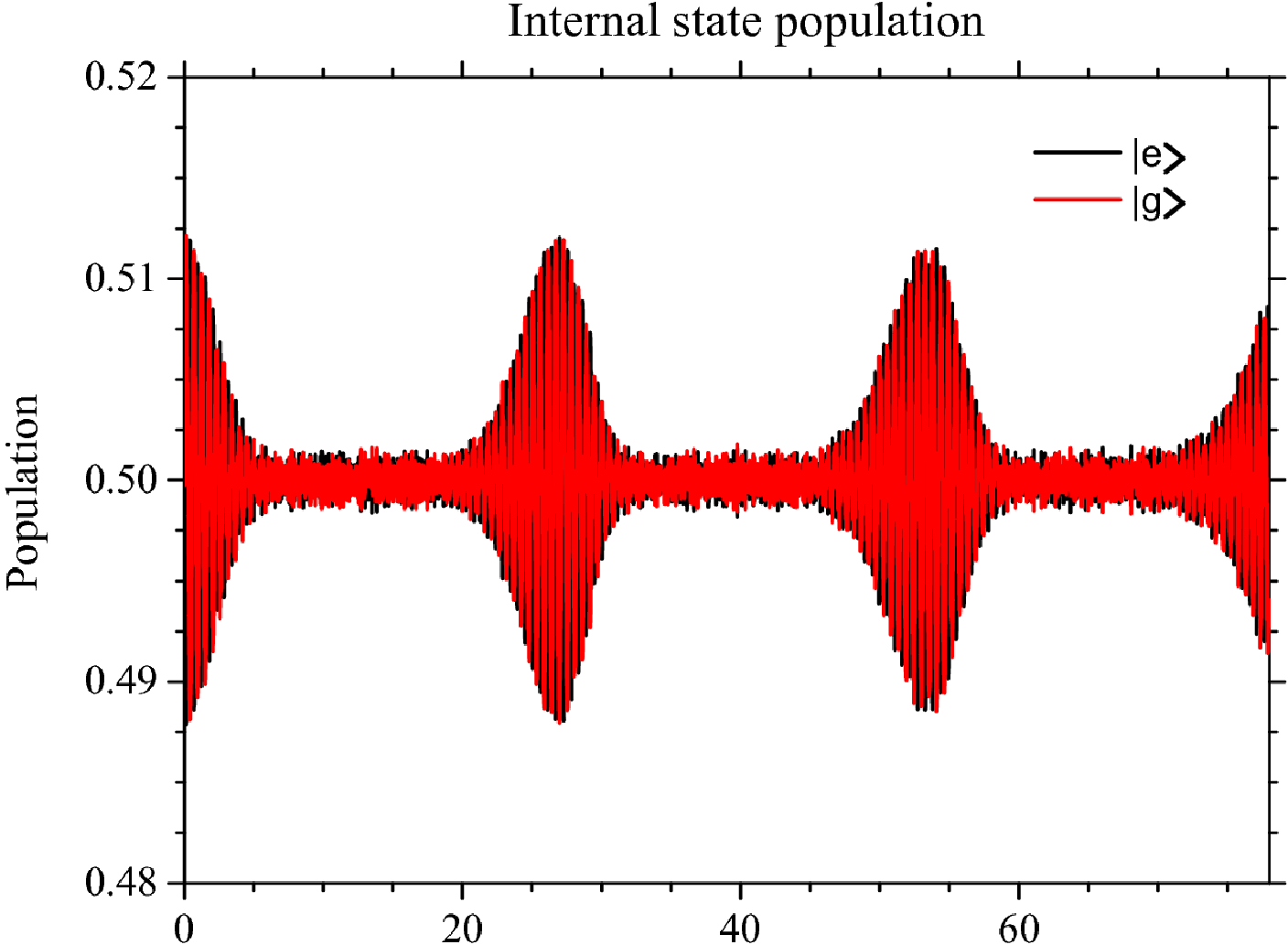}\\[1ex]
  \raisebox{3.5cm}{\small (c)}\includegraphics[width=7cm]{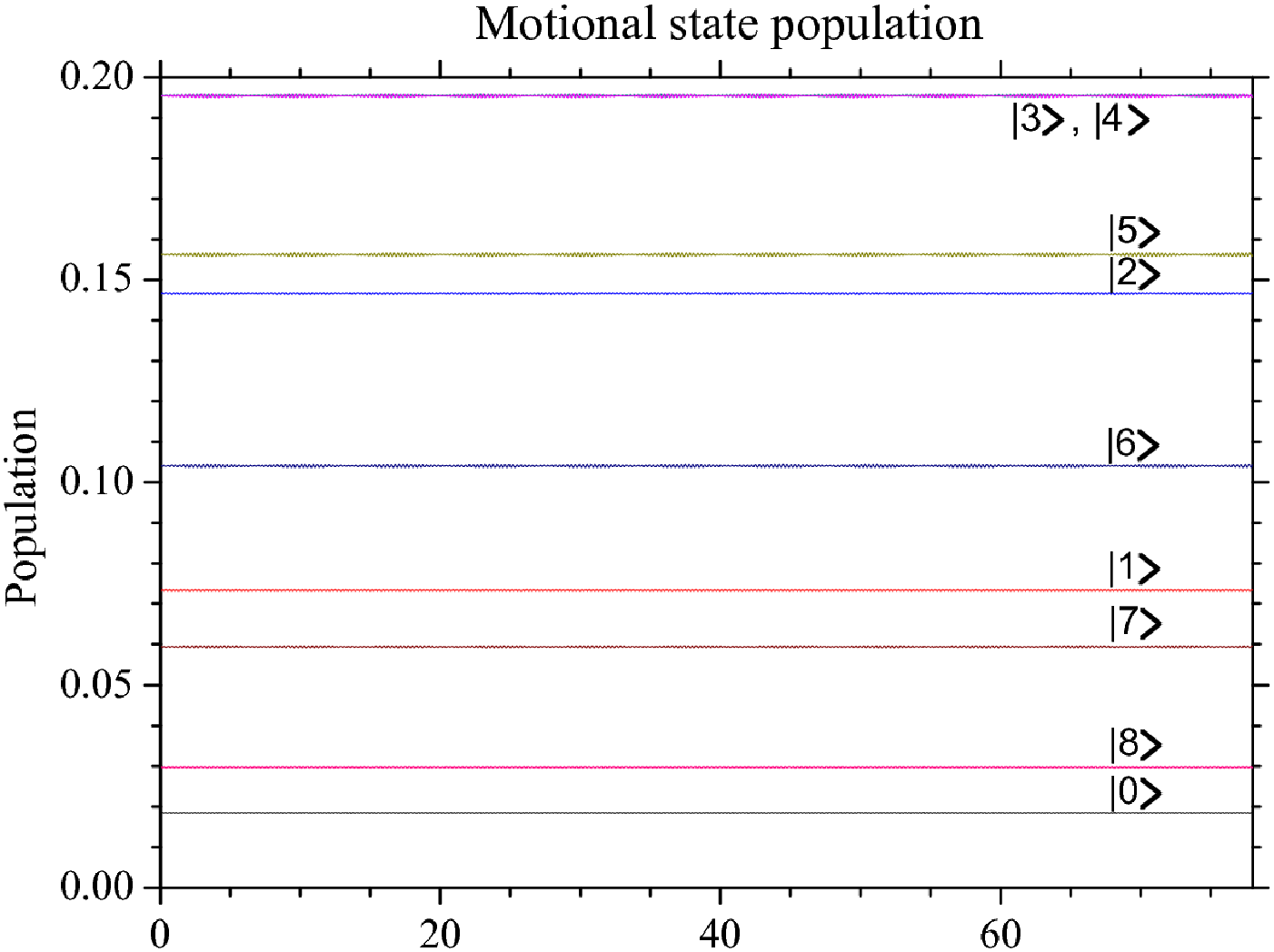}
\end{center}
\caption{(a) Evolution of the first moments of the oscillator position 
  $X$ and momentum $Y$, starting from the coherent state $|\alpha=2\rangle$, for 
  $\eta=0.2$, $\Omega=2\pi\times100$~kHz, and $\omega=2\pi\times 3$~MHz.
  Compare to fig.~\ref{kerrideal} (line colors and styles follow the same 
  convention). Physical time is displayed on the lower axis, whilst 
  the effective Kerr time appears on the upper axis. The fidelity of the 
  motional state evolution as compared to the ideal Kerr state 
  of eq.~\ref{eq:kerr} is depicted by the gray curve (right axis). 
  (b) Internal state population. (c) Phonon population.}
\label{evolution}
\end{figure}

An example of pulse operation as a function of time for $\eta=0.2$ is shown 
in fig.~\ref{evolution}.  For this particular choice of
parameters, the cat state was obtained in 13~ms interaction time with
98.5\% fidelity. More complex states were generated faster with higher
fidelity. The fast ripples on the internal state population
(fig.~\ref{evolution}b) reach maximum amplitudes of 1\% (although
typically 0.1\%), whilst the phonon state populations
(fig.~\ref{evolution}c) fluctuates within 0.1\%. Those off-resonant
excitations, however, do not accumulate dramatically over time.
Those results confirm that the dynamics using higher-order terms of the 
Hamiltonian does not hinder in itself the attainability of high fidelities.

\begin{figure}
\centering
\includegraphics[scale=0.34]{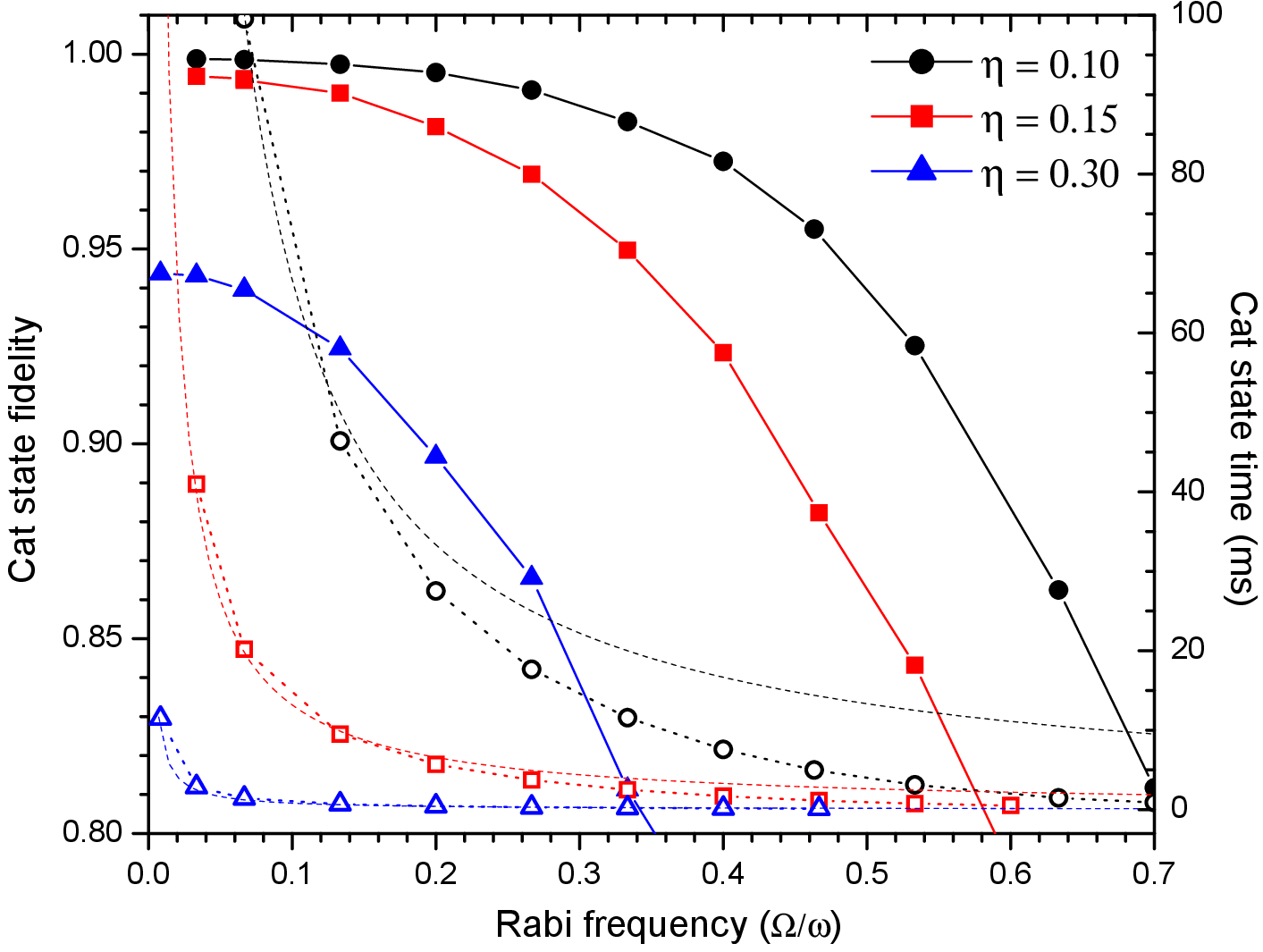}
\caption{Cat state fidelity as a function of $\Omega$, for three 
  different values of $\eta$ (full symbols). The times needed to 
  produce the cat state are displayed on the right axis (open symbols). 
  The dashed lines result from the approximated effective evolution 
  time of eq.~\ref{taueff}. Each point results from a numerical 
  calculation of the quantum evolution.}
\label{fidtimeXrabi}
\end{figure}

The cat state fidelity as a function of Rabi frequency 
for three values of $\eta$ is depicted on fig.~\ref{fidtimeXrabi}. The smaller values of $\eta$, 
the higher fidelities for the cat state. The physical time necessary 
for the formation of the cat state is shown on the right axis.  In principle, 
fidelity of the state can be arbitrarily close to unity although the time required for its creation 
increases. In practice, decoherence will limit the maximum operation time. Thus, the 
optimum pulse duration will depend on the particular experimental conditions. 
The tradeoff between increasing Rabi frequency and decreasing the ideal fidelity results 
in shorter pulse duration. In fact, the operation could be faster even for low 
values of $\eta$ if the Rabi frequency would be increased to values comparable to the oscillator frequency. 
It is important to note however, that eq.~(\ref{taueff}) is no longer valid in the limit of 
small $\eta$ and large $\Omega$ due to saturation effects and off-resonant excitations. 

\begin{figure}
\centering
\includegraphics[scale=0.34]{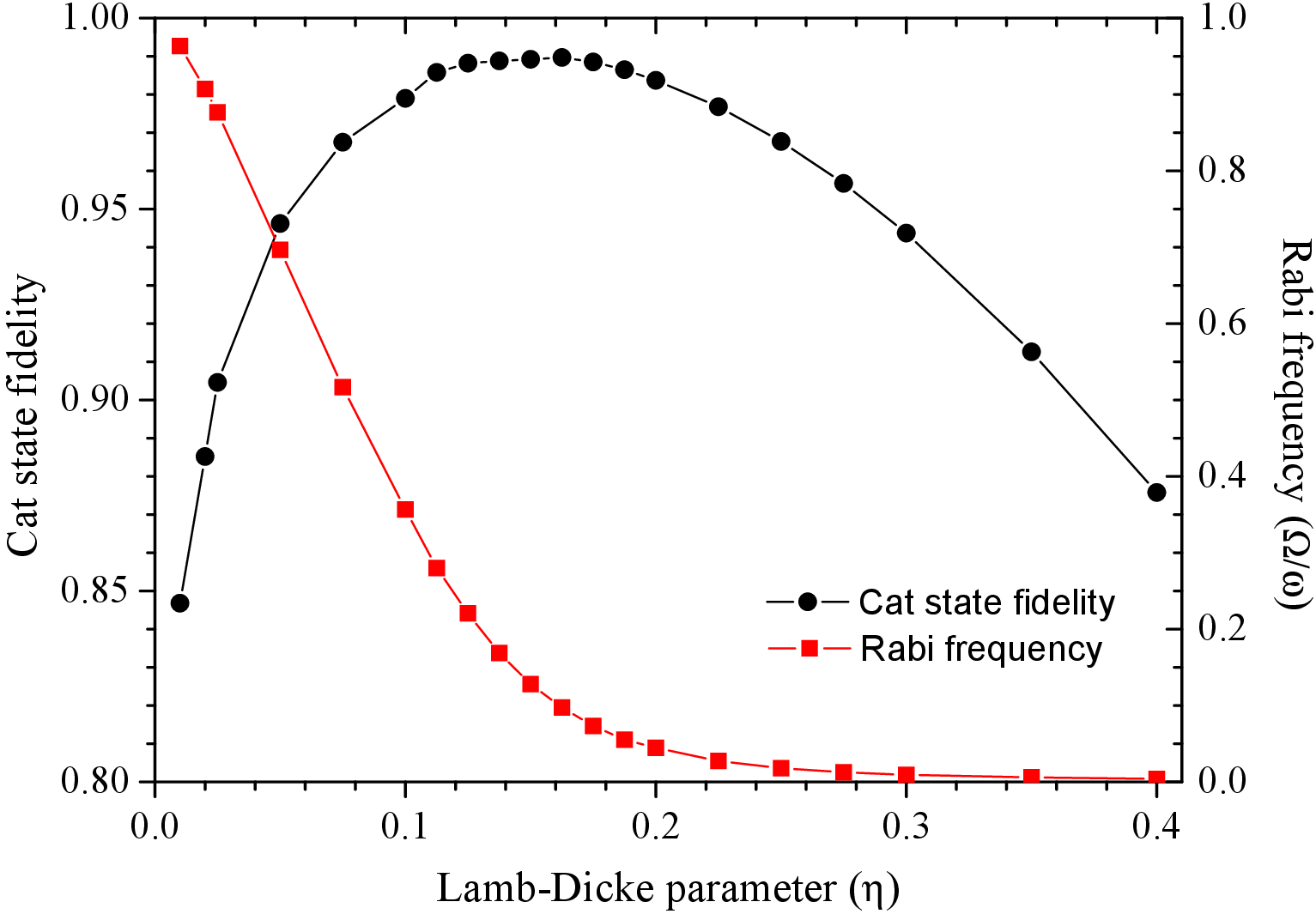}
\caption{Cat state fidelity as a function of the Lamb-Dicke parameter 
  $\eta$ for a fixed cat state creation time of 10~ms (black circles). 
  The required Rabi frequencies relative to the oscillator frequency 
  ($\Omega/\omega$) are indicated by the red squares. Each point 
  results from a numerical calculation of the quantum evolution.}
\label{cat10ms}
\end{figure}

The more realistic experimental situation where the cat state generation has to be 
completed given a fixed interaction time was also examined. The cat state fidelity 
as a function of the Lamb-Dicke parameter $\eta$ for a 10~ms pulse duration is 
shown in fig.~\ref{cat10ms}. Rabi frequencies are indicated on the right
axis of the figure. The optimum cat state fidelity of 99\% was found for
$\eta\approx 0.15$ for our choice of parameters. The smaller $\eta$ is, the lower 
the fidelity becomes due to the large values of $\Omega$ which drive undesired transitions. 
As expected, the higher order terms in eq.~(\ref{hamiltonian}) contribute significantly 
for larger $\eta$ and hinder the ideal Kerr dynamics.

\begin{figure}
\centering
\includegraphics[scale=0.34]{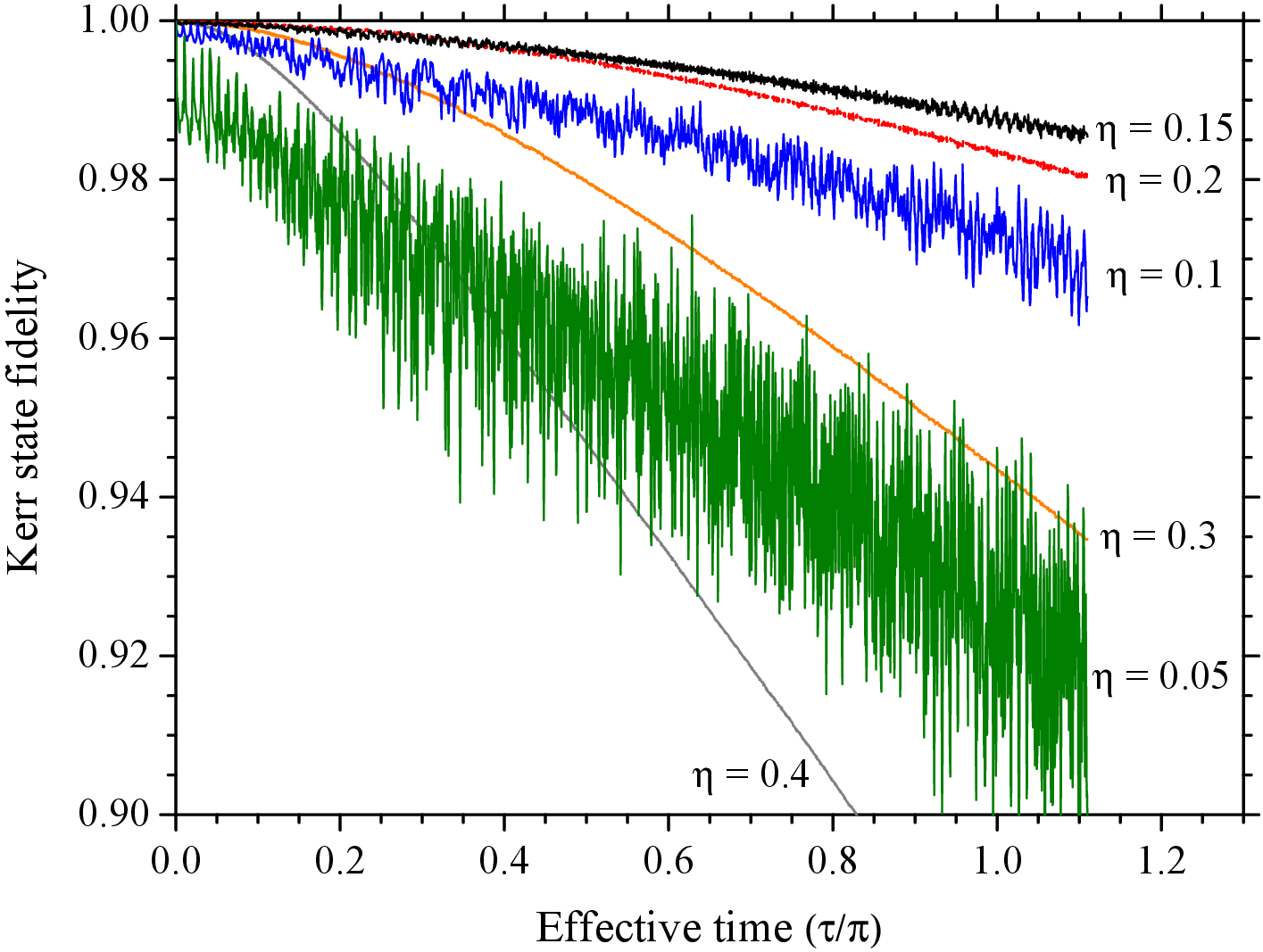}
\caption{Fidelity of the motional state evolution as compared to the 
  ideal Kerr state. A fixed time of 10~ms (corresponding to $\tau=\pi$) 
  is imposed for the formation of the cat state (fig.~\ref{cat10ms}). }
\label{timefid}
\end{figure}

%Although our method allows to generate a small and a big cat state in the same fixed time, 
%the bigger cat states will decohere faster. Coupling to a thermal reservoir is expected to
%result in the loss of visibility in the interference pattern $V=\exp(-|\alpha|^2 \gamma t )$, 
%where $\gamma$ is the coupling constant and $t$ the coupling time~\cite{Leibfried2003}. We estimated that for 
%a heating rate $\gamma=3$ quanta per second and time $t=6ms$, $V=0.99$ is obtained for $\alpha <0.7$ and 
%$V=0.9$ for $\alpha <2.4$. For time $t=10ms$, $V=0.99$ is achieved for $\alpha <0.6$, whereas 
%$V=0.9$ for $\alpha <1.9$. However, we emphasis that the time for a cat state generation is 
%the longest one. All other quantum superpositions are generated faster and thus, larger coherent 
%state amplitudes are possible.

Although the Kerr Hamiltonian allows one to generate either `small' or `big' 
cats using the same interaction time, bigger cats will decohere faster, 
since coupling to a thermal reservoir results in loss of visibility of the 
interference pattern in phase space (with exponential dependence on the cat size)~\cite{Leibfried2003}. 
However, the cat state is the final product of the Kerr interaction, and as
such takes the longest time to be created. Other interesting macroscopic quantum 
superpositions could therefore coherently occupy larger areas in phase space 
albeit for shorter durations.

Comparison between the quantum evolution of some of the states
presented in fig.~\ref{cat10ms} and the ideal Kerr states is presented in 
fig.~\ref{timefid}. For the optimum $\eta$, a superposition of four coherent
states is produced with 99.6\% fidelity, whilst three superposed
coherent states can be obtained with 99.3\% fidelity. The first
negativities of the Wigner function are observed already after
500~$\mu$s interaction time, and dramatic effects in sub-Planck
regions of phase space appear after 2~ms with more than 99.9\%
fidelity.  We stress that these nearly ideal fidelities indicate that the actual 
experimental limitations will not be a consequence of the form of the
Hamiltonian, but rather decoherence and imperfections which will
inherently depend on specific details of the experimental setup.
For instance, larger cats can be expected to experience stronger
decoherence~\cite{haroche}, effectively requiring faster nonlinear 
dynamics.

From the perspective of the interaction Hamiltonian, the upper limit 
on the fidelity for a fixed cat state creation
time results either from off-resonant excitation involving 
the phonon sidebands or from higher-order terms in the 
Hamiltonian. Off-resonant excitations cause noisier evolution
for smaller $\eta$, while for larger values of $\eta$ the evolution 
becomes smoother (fig.~\ref{timefid}). These excitations can be thus avoided 
by increasing the oscillator frequency. Taking all necessary 
compensations into account, our calculations indicate that a 
higher oscillator frequency has a positive effect on the state 
fidelity.

\section{Conclusion}

We have presented a detailed investigation of a practical method of
involving Kerr non-Gaussian state generation in the vibrational mode of trapped 
ions~\cite{davidovichkerr}. The method employs only one laser pulse, 
directly connecting an initial coherent state to the desired 
highly nonclassical Kerr state. Since only a resonant carrier 
pulse is needed to harness the natural Kerr nonlinearity of 
trapped ions, this scheme can be performed even in
very simple setups, such as the stylus trap~\cite{stylus}. 
An optimized setup would be desirable, however, where the 
trap size should be large enough to avoid motional heating
and decoherence at the same time that a high oscillator frequency 
would hinder off-resonant excitations even for fast dynamics.

Special attention has been paid to the cat states, for their importance
in quantum information applications. However, other complex quantum
states such as superpositions of several coherent states can also be 
generated in the process. Highly non-Gaussian states with negative values
on the Wigner function are obtained for very short interaction times.

The scheme produces large superpositions in phase space 
using the same time resources necessary for small superpositions, 
without the need for artificially truncating the population in the 
Fock state basis. 

\acknowledgments

We acknowledge support from the Alexander von Humboldt
foundation. M. S.  acknowledges further support from the MNiSW Grant
No. 2619/B/H03/2010/38.  M. S. thanks I. Cirac for discussions.

\end{document}